# Multivariate Gaussian Process Incorporated Predictive Model for Stream Turbine Power Plant

*Prama Debnath*
*Department of Computer Engineering, American International University-Bangladesh, Dhaka, Bangladesh*

*Mithun Ghosh\**
*Department of Systems and Industrial Engineering, University of Arizona, Tucson, USA.*

**Abstract:**
Steam power turbine-based power plant approximately contributes 90% of the total electricity produced in the United States. Mainly steam turbine consists of multiple types of turbine, boiler, attemperator, reheater, etc. Power is produced through the steam with high pressure and temperature that is conducted by the turbines. The total power generation of the power plant is highly nonlinear considering all these elements in the model. We perform a predictive modeling approach to detect the power generation from these turbines by the Gaussian process (GP) model. As there are multiple interconnected turbines, we consider a multivariate Gaussian process (MGP) modeling to predict the power generation from these turbines which can capture the cross-correlations between the turbines. Also, the sensitivity analysis of the input parameters is constructed for each turbine to find out the most important parameters.

**Introduction:**
Usually, a steam turbine power plant can be classified according to the source of its steam generations. For example, that source can be coal, solar power, or nuclear. A steam cycle power plant is usually maintaining the Rankine cycle. The cycle begins with the generation of steam from the water by flowing through the boiler. Then, this steam rotated the turbine shafts when it goes through it which in turn creates electricity. In order to recycle the steam, it is allowed to go through the condenser and converted back to the water. Later a cycle begins with the water pumped back into the boiler.

During the Ranking cycle, the steam is initially cooled down into the liquid water through the condenser and then heated back to the steam through the boiler.  Due to the energy efficiency of the liquid compared to the steam, this conversion is required. Also, compared to a compressor, a pump works more efficiently with less amount of energy. In order to improve the cycle efficiency, there may be extra parts that are added based on the requirement, for example, reheater, feedwater heaters, moisture separators, etc.

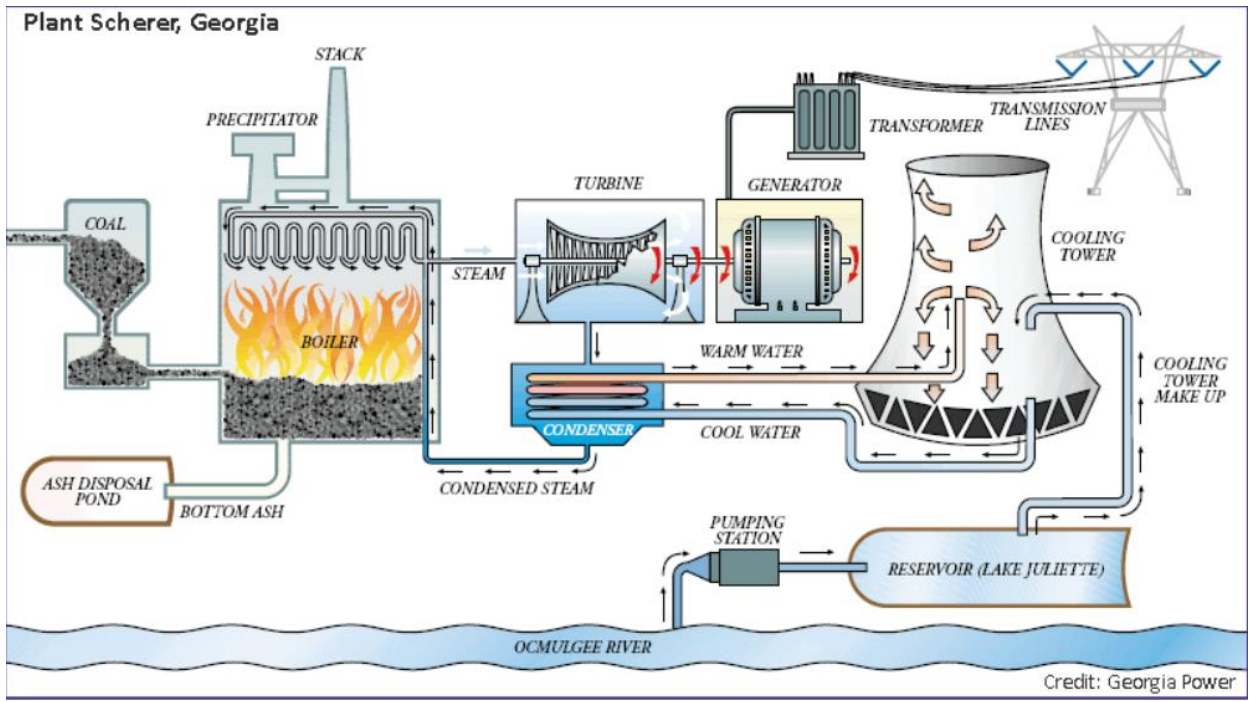

Figure 01: A brief overview of a coal-based steam turbine.

To achieve the full efficiency of the steam, a reheater is used to reheat the steam along its way back to the boiler and subsequently lower pressure turbine. Although this requires some extra additional cost with the expense of extra power gain from the turbine. To reduce the damage in the turbine blades by moistures, a moisture remover is often used. Feedwater heaters act as a heat exchanger which can be an open or closed design. The functionality of the feedwater is to act as a heater to heat the portion of post-condenser water steam coming from the high-pressure turbine and sent it to the boiler. This helps to reduce the needed heat by the boiler.

A schematic view of the coal-based steam turbine of the Georgia power plant is given in Figure 01 that is taken from [1]. These turbines from Figure 1 are named after the incoming pressure of the steam and they are interconnected in series. Figure 2 shows an enlarge details view of these turbines that are named high pressure (HPT), intermediate pressure (IPT), and low pressure (LPT) turbines. As mentioned in their name, these turbines are different in generating power because of different pressure, temperature intakes, and also their internal structure.

Thus, a steam turbine is composed of many different complex parts and to get the optimal operating conditions, the whole systems need to be considered. But this becomes unrealistic as most of the parts do not have a real-world metrics to measures. We considered all the aspects of a power plant to generate the data. While for the modeling part, we used the turbine section as it the ultimately generating the output power. The relationship between different components in the powerplant is highly nonlinear. Also, a simple model is not enough to capture the nonlinearity and also the correlations between turbines. We developed a multivariate and multitask Gaussian Process model that can handle the nonlinearity between the features and the outputs, also provided us with a prediction band to compute the uncertainty in the model.

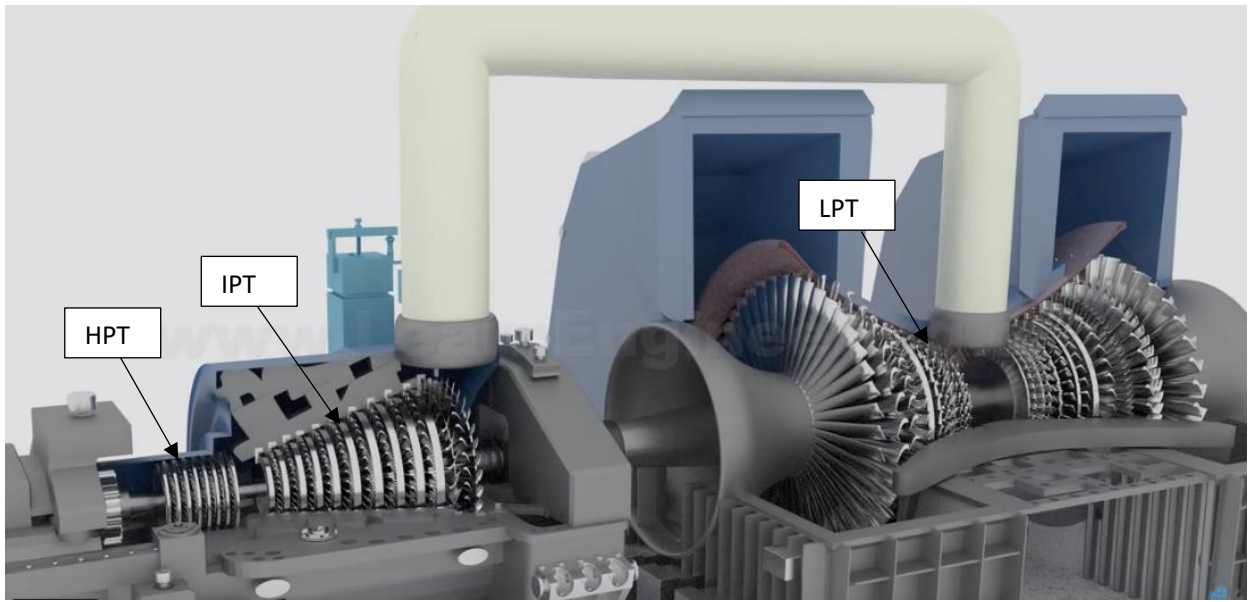

Figure 02: Turbines connected in series.

The dynamics of steam turbines are constructed by a few models [3-5]. Most of these models ignored the other important aspects of turbines and do a simple mapping from input to output without considering the important complex relationship between and within them. To improve the performance in terms of control theory, we need a complex model that is not based on some infeasible conditions. After all, we need a satisfactory level of precision to model the output of a powerplant [7].

Apart from the control theory, another popular technique is the identification technique which is based on mathematical models. Data is feed to the model to observe the real-time system performance. The purpose of the model is to find any abnormalities during the normal operation without any external excitation. There is also a limitation in these models as it is not possible all time to get the flow of the operating data reliably without external excitation [8]. There is no well-established parametric model that can capture the input-output relationship well enough and which can be used in the power plant.

Genetic algorithms (GA) have become very popular recently due to its ability to seek a global solution to a multimodal loss function. It is preferable to find the optimal parameter where we have a large amount of nonlinearity with complicated structures compared to the conventional methods [9]. Genetic algorithms are better suited for the optimization of factor vectors. However, if we try to optimize a vector of continuous variables, we will face a tremendous loss of performance.

Gaussian process (GP) or kriging is one of the most popular of today's machine learning advancement. GP is a very popular modeling tool in the machine learning community due to its flexibility to capture nonlinearities, quantify correlations, and uncertainty measuremnts. It is used to capture the spatial or temporal relationship based on some kernel functions. GP is nonparametric in nature and also calls the best linear unbiased estimator. Multitask Gaussian Process (MGP) is the direct extension of the GP, where we take consideration of more than one output. MGP has a cross-correlation structure that can capture both within and between task correlations in the modeling. The nonstationary version of the MGP can also be found in [16]. Bayesian approach of the modeling is also demonstrated in [17].

Likewise, in our steam turbine model, we have three types of the turbine and thus we get three-task GP. The turbine correlations are constructed by MGP. MGP helps to capture more information from the data than univariate GP by lending information from the other levels. Thus, we will use this transfer learning framework to improve our model accuracy. We intend to show the performance of MGP with GP.

In section 2, we developed the Simulink model for the analysis of the transient response of steam turbine subsections based on the mathematical models, thermodynamic state conversion, and semi-empirical equations developed in [10]. Then, the simulation data is generated from these Simulink models of the turbine. For the design point generation, some design of experiment criteria is being followed. In section 3, the simulation model is compared with the mathematical model. In section 4, the conclusion and future direction of the work are discussed.

## 2. Modeling:

### 2.1 Simulink Modelling:

The lack of real data availability of a powerplant has given us the idea of the analysis to be based on the simulated data. To build a Simulink model, we need to develop the condition and the operating criteria for each element of the power plant. While this task requires a lot of labor, a Simulink model consisting of three turbines is built based on the literature reviews. The overall Simulink structure of the powerplant is given below in Figure 03. The details of these turbines are also structured in the subsequent plots. Their Simulink modeling schema is also described in Figure 04, 05, and 06.

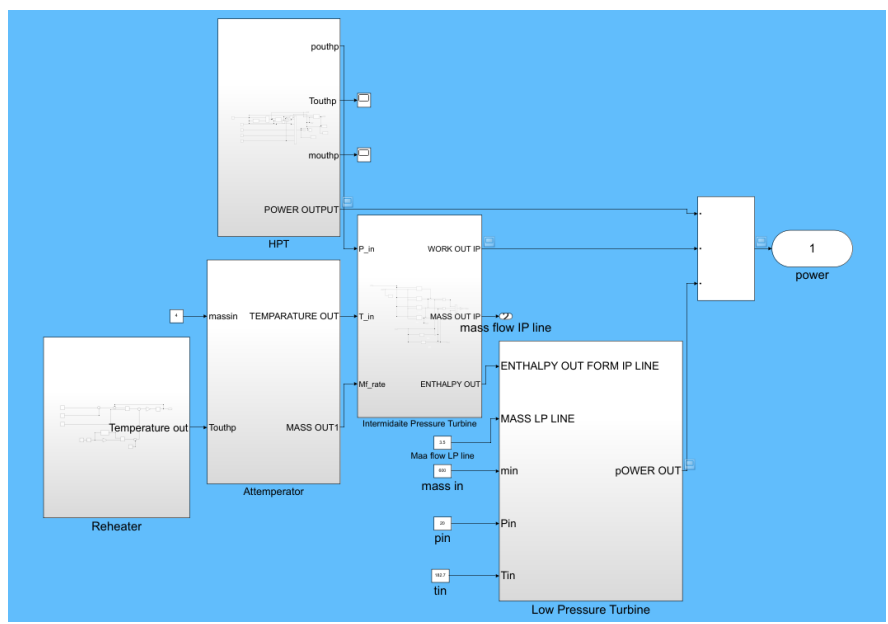

Figure 03: The Simulink model of the Power Plant

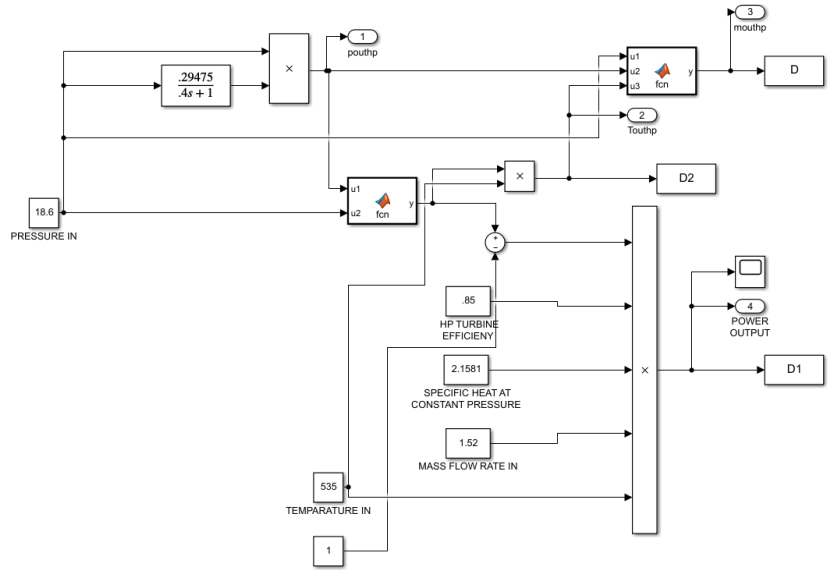

Figure 04: The Simulink model of the High pressure turbine (HPT)

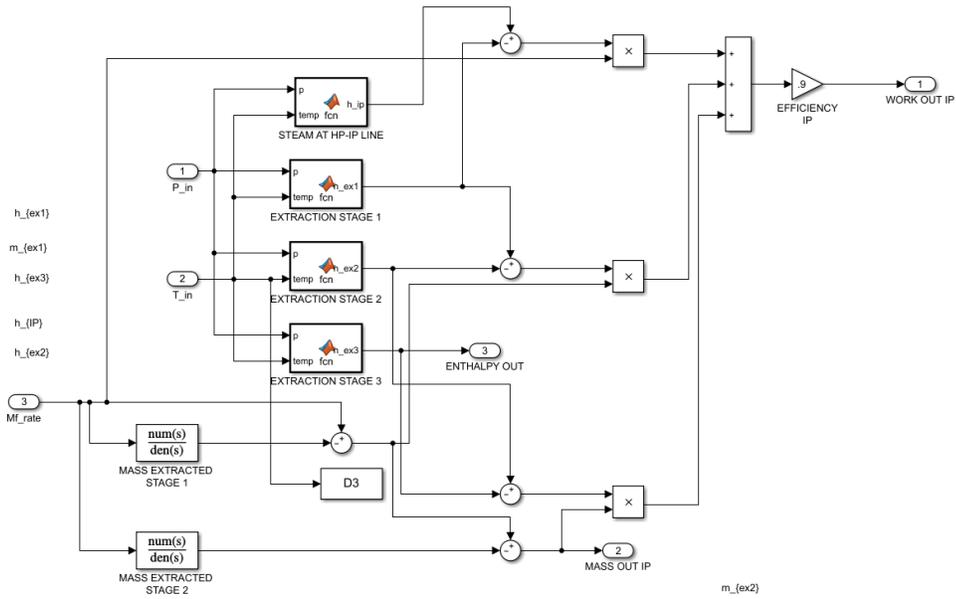

Figure 05: The Simulink model of the Intermediate pressure turbine (IPT)

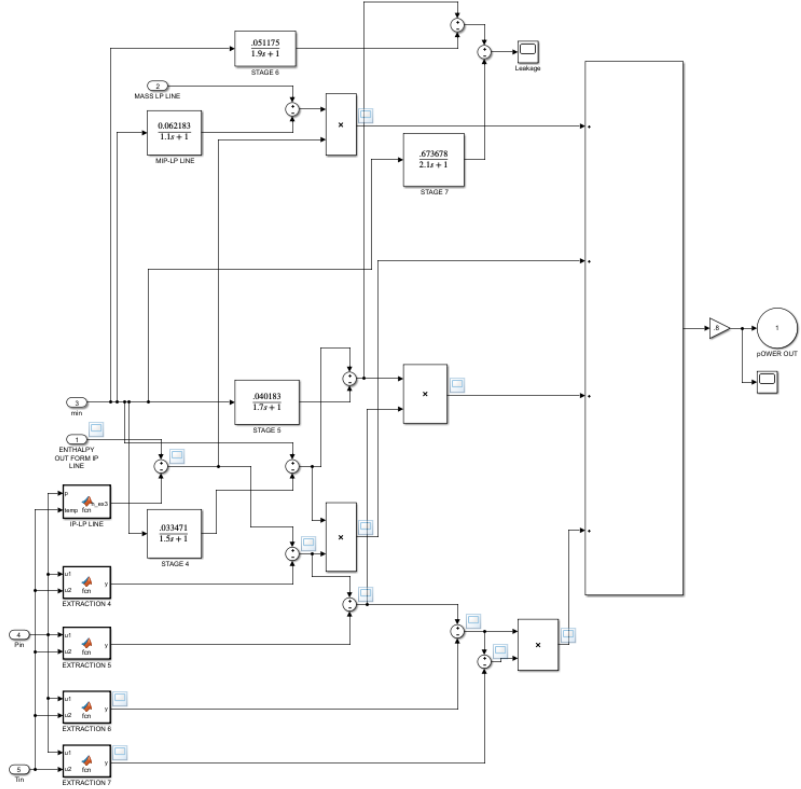

Figure 06: The Simulink model of the Low pressure turbine (LPT)

## 2.2 Mathematical modeling:

Let's assume that we have a $k$ number of turbines with $p$ independently observed sample runs. The $i$th ($i = 1,2,\ldots,K$) turbine on the $m$th sample run ($m = 1,2,\ldots,M$) has $N_i$ observations defined over $l$ explanatory variables. So, the measurement of the $i$th turbine at the $j$th combination of variable ($j = 1,2,\ldots N_i$) of the $m$th sample can be formulated as the following model structure:

$$\mathbf{y}(\mathbf{x}) = \begin{bmatrix} y_1(\mathbf{x}) \\ y_2(\mathbf{x}) \\ \vdots \\ y_K(\mathbf{x}) \end{bmatrix} = \begin{bmatrix} \mathbf{f}_1(\mathbf{x})^T \boldsymbol{\beta}_1 \\ \mathbf{f}_2(\mathbf{x})^T \boldsymbol{\beta}_2 \\ \vdots \\ \mathbf{f}_K(\mathbf{x})^T \boldsymbol{\beta}_K \end{bmatrix} + \begin{bmatrix} \varepsilon_1(\mathbf{x}) \\ \varepsilon_2(\mathbf{x}) \\ \vdots \\ \varepsilon_K(\mathbf{x}) \end{bmatrix}$$

$$= \mathbf{F}(\mathbf{x})^T \boldsymbol{\beta} + \boldsymbol{\varepsilon}(\mathbf{x}), \tag{1}$$

where $\mathbf{f}_i(\mathbf{x}) = [f_{i1}(\mathbf{x}), f_{i2}(\mathbf{x}), \ldots, f_{iq_i}(\mathbf{x})]^T$ is the regression function of any order and $\boldsymbol{\beta}_i = [\beta_{i1}, \beta_{i2}, \ldots, \beta_{iq_i}]^T$ are corresponding coefficient. $\mathbf{f}_i(\cdot)^T \boldsymbol{\beta}_i$ formed the fixed part of the model (1). $\varepsilon_i$ is the measurement error for the $i$th turbine in the $m$th sample, respectively. Assuming iid measurement error, $\varepsilon_i(\cdot) \sim N(0, \sigma_\varepsilon^2)$, which can be easily relaxed.

The error term $\varepsilon(x)$ is considered stationary, with the following covariance function $C_{i,j}(\mathbf{x}, \mathbf{x}') = cov(\varepsilon_i(\mathbf{x}), \varepsilon_j(\mathbf{x}')) = \sigma_i \sigma_j R_{i,j}(\mathbf{x}, \mathbf{x}')$, where $R_{i,j}(\mathbf{x}, \mathbf{x}')$ is the correlation between $\mathbf{x}$ of the $i$th turbine and $\mathbf{x}'$ of the $j$th turbine, $\sigma_i^2$ and $\sigma_j^2$ are variance in $i$th and $j$th turbine's power generation, respectively. The formulation of an MGP as in Eq. (1) is quite general as it is directly extended from a standard univariate GP model framework, which it reduces to GP if $K = 1$.

We can consider a single univariate GP for each turbine output which will ignore the cross-correlations between them. Whereas in the MGP all the turbines are considered in a single model. We used the nonseparable covariance function, which is much more flexible compared to the flexible one, blends the information of different turbines by sharing information across the. There are mainly many ways to build valid nonseparable cross-covariance functions, such as: coregionalization and kernel convolution. A flexible cross-correlation structure was proposed by [11]. In this article, considering the flexibility, we adopted their suggested covariance:

$$R(\mathbf{x}_i, \mathbf{x}_j) = \sigma_i \sigma_j T_{ij} \frac{\exp\{-(\mathbf{x}_i - \mathbf{x}_j)^T (\Phi_i^{-1}/2 + \Phi_j^{-1}/2)^{-1}((\mathbf{x}_i - \mathbf{x}_j)\}}{|(\Phi_i/2 + \Phi_j/2)(\Phi_i^{-1}/2 + \Phi_j^{-1}/2)|^{1/4}},$$

where $\varphi_i = \text{diag}(\varphi_{i,1}, \varphi_{i,2}, \ldots, \varphi_{i,p})$, represents the roughness parameters of the covariance, $T_{i,j} \in [-1, 1]$ is the ($i$th, $j$th) element of a $K \times K$ matrix which represents the cross correlation between two turbines ($i$th, $j$th) with $T_{i,i} = 1$. **T** must be a positive definite with unit diagonal elements (PDUDE). Hypersphere Decomposition (HD) is used to construct the **T** matrix which helps it to produce a valid cross correlation.

HD is mainly introduced to construct a valid positive definite matrix with unit diagonal element (PDUDE) by [12-13]. At first, T is decomposed into Cholesky decomposition such that:

$$\mathbf{T} = \mathbf{EE}',$$

where $\mathbf{E} = \{E_{rs}\}$ is a $k \times k$ lower triangular matrix with positive diagonals. The elements of **E** are then formulated in the spherical coordinate systems in the following way:

$$E_{rs} = \begin{cases} 1, & r = s = 1 \\ \cos(\omega_{rs}), & 1 \leq r < K, s = 1 \\ \cos(\omega_{rs}) \prod_{q=1}^{s-1} \sin(\omega_{rt}), & 1 \leq r < K, 1 < s < r \\ \prod_{q=1}^{s-1} \sin(\omega_{rt}), & 1 \leq r < K, s = r, \end{cases}$$

where $\omega_{rs} \in (0, \pi)$ which is collectively denoted by $\Omega = \omega_{rs}\{r > s\}$. Thus, matrix **T** has to be constructed in a sequential manner. Because $\omega_{rs} \in (0, \pi)$, which are the entries in **T**, are in between [-1,1] which represents correlation across different profiles. The mapping here is one to one between PDUDE matrix and **ω** and for any **ω** we will always get PDUDE. The number of elements in matrix $\mathbf{T}$ is $\frac{K(K-1)}{2}$ for a $K$ variate turbine. Thus, with $\mathbf{T}$ we are constructing cross correlation or between turbine correlation.

## *2.3 Factor screening and optimization procedure:*

We use L1 regularization to the fixed term to screen out any factor that is not contributing in the model. We add this penalty term into the log-likelihood function where we estimate the parameter value. Factors with less influence will be forced to have no influence by producing a sparse solution.

For the estimation of the parameter, we maximized the log likelihood function. Considering the observation in the $m$th sample of the levels, $\mathbf{y} = [\mathbf{y}_1; \ldots; \mathbf{y}_k]$, then the penalized log-likelihood function which is to be maximized:

$$\log L(\boldsymbol{\beta}, \boldsymbol{\sigma}, \boldsymbol{\Phi}, \mathbf{T}, \sigma_\varepsilon^2; \mathbf{y}_1, \ldots, \mathbf{y}_k)) - \lambda|\boldsymbol{\beta}|$$
$$= -\frac{1}{2}(\log |\mathbf{R}| + (\mathbf{y} - \mathbf{F}\boldsymbol{\beta})^T \mathbf{R}^{-1}(\mathbf{y} - \mathbf{F}\boldsymbol{\beta})) - \lambda|\boldsymbol{\beta}| + \text{constant}$$

where $\mathbf{F} = \text{blkdiag}(\mathbf{f}_1, \ldots, \mathbf{f}_k)$ with dimension of $N_i \times q_i$ of $i$ th block being $\mathbf{f}_i = [\mathbf{f}_{(i,1)}, \ldots, \mathbf{f}_{(i,N_i)}]^T$. $\mathbf{R}$ is the covariance matrix. $\boldsymbol{\Phi} = [\Phi_1; \ldots; \Phi_k]$ are the roughness parameters. $\lambda$ is a penalty term. High values of $\lambda$ drag most of the coefficients, $\boldsymbol{\beta}$ to zeros which implies that there are not many influences of the respective $\mathbf{x}$'s and vice versa for very high value of $\lambda$. All the parameters are derived by maximizing the log likelihood function.

## *2.4 Prediction:*

After tuning the parameters of the model, we can get the prediction for an untried input. The prediction equation for the MGP model at the new input point $x_0$ is:

$$\hat{y}(\mathbf{x}_0) = \mathbf{f}(\mathbf{x}_0)^T \widehat{\boldsymbol{\beta}} + \mathbf{r}^T(\mathbf{x}_0)\mathbf{R}^{-1}(\mathbf{y} - \mathbf{F}\widehat{\boldsymbol{\beta}}),$$

where $\mathbf{r}(\mathbf{x}_0) = \mathbf{R}(\mathbf{x}_0, \mathbf{x})$.

## *2.5 Design of Experiment:*

A space-filling design criterion is used to collect the observation from the Simulink model. Latin Hypercube Sampling (LHS), which is one of the most popular sampling approaches, is used to generate the trivariate training dataset for the power plant model for 50 observations with 30 replications. LHS is the most preferable random sampling approach for high dimensional features to construct computer experiments or for Monte-Carlo integration. We use six-dimensional LHS with minimax criteria to ensure the exploration of most of the area in space for the input designs. Similarly, another 50 observations and 30 replications are generated for the test dataset. From the literature, the selected six input parameters that contribute most in the power plant and their selected range for the simulation is described below:

Table 01: Inputs with range

| Variable | Input | Range |
|---|---|---|
| 1 | Pressure | 35-10 MPa |
| 2 | Temperature | 2000-500 K |
| 3 | Mass flow | 3-2.2 kg/s |
| 4 | Electric grid frequency | 60-50 Hz |
| 5 | Number of blades | 20-5 |
| 6 | Boiler temperature | 650-550 K |

As this is a dynamic model, we will get different output in the simulation at the different time step. But at a certain time, output reaches steady states when its values do not change with time. We record that value for each simulation and treat it as the final output value in the modeling.

**3. Results:**

We fit our model with the Univariate or Independent GP and the Multivariate GP framework. The RMSE value for each model is described in Figure 07. Both of the models are in close range with low RMSE which proves the predictive capability of GP. The MGP model outranked the GP model slightly by capturing the cross correlation correctly between the turbines. This gives the MGP model edge over the Univariate GP model. If we can analyze the cross-correlation parameters in Table 02, it seems the HPT and IPT are correlated strongly with the value of 0.752. Whereas, the correlation of other turbines is low but positive. It is as expected. With the increase in the steam temperature and pressure values, these correlation parameters should increase. If there were not pressure and temperature loss, these turbines would have the ideal cross-correlation of 1.

Table 02: The cross-correlation structure of the turbines

| 1 | 0.7592 | 0.0575 |
|---|---|---|
| 0.7592 | 1 | 0.0453 |
| 0.0575 | 0.0453 | 1 |

We also perform the sensitivity analysis of the MGP model by Elementary Effect (EE) method. EE method quantifies the noninfluential inputs in the computationally costly models. It ranks the inputs in order of importance and helps to discard the less important one when we have a large number of inputs. The details can be found in [14] and [15].

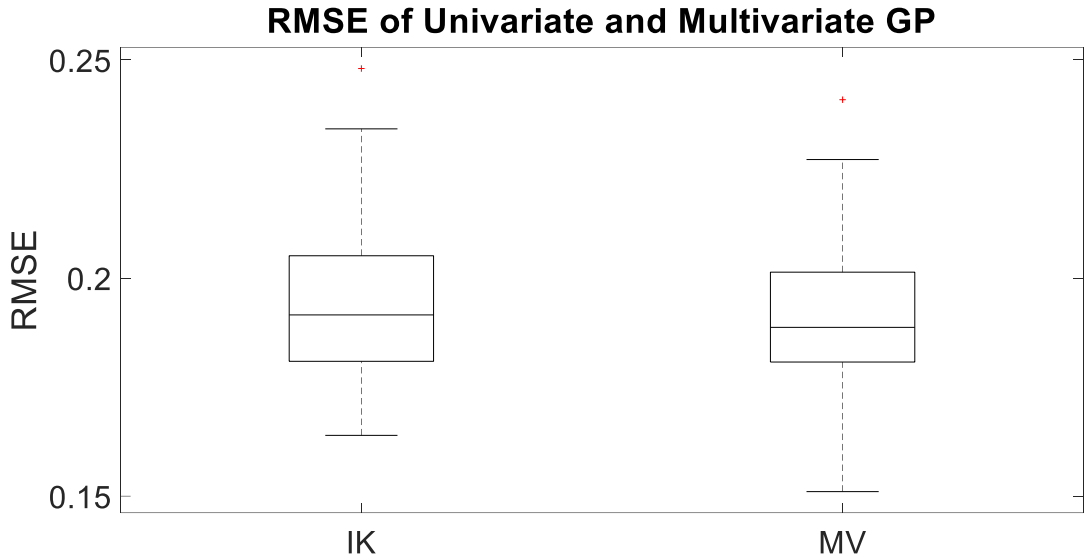

Figure 07. Two random realization of the turbine model with actual and predicted value.

A for 10 tours and 30% in variables of its range is used for the analysis. In Figure 08 the sensitivity of the three turbines is described where we can see the plot of mean vs variance of the parameters. The non-influent inputs can be categorized based on the low values of both mean and variance. Variables 6 have too large mean and variance which is why it is excluded in the plot. Most of the variables in the turbine model has high mean and variance which redefine the high nonlinearity of the turbine model in its input and output relations. Especially the variable 1,5 and 6 which are pressure, number of the blades and boiler temperature are more influential than others.

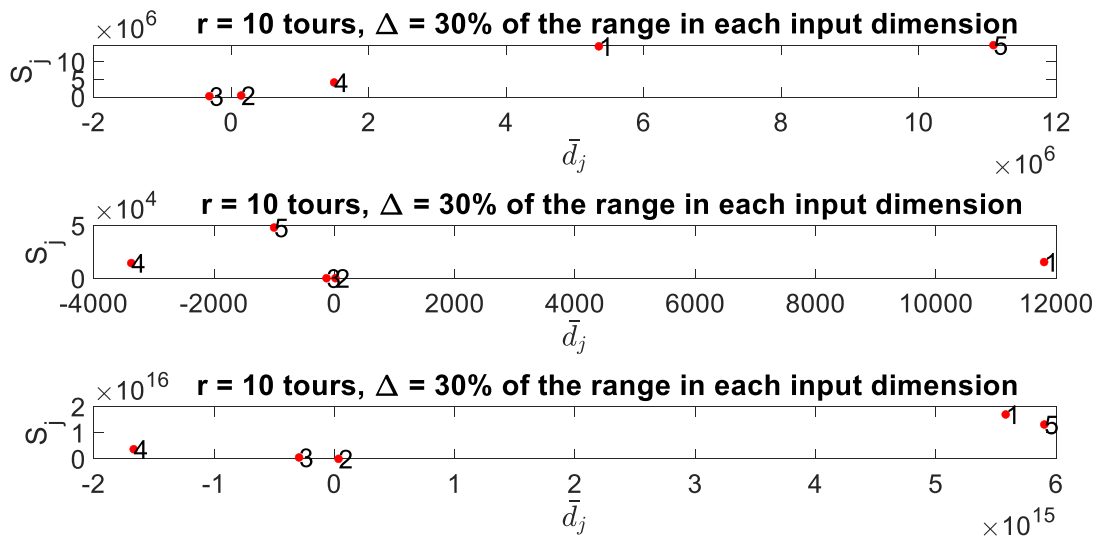

Figure 08: Sensitivity analysis of the turbine model by EE algorithm. From the top, the plots are for HPT, IPT and LPT, correspondingly.

## 4. Conclusion

A Power plant model is highly nonlinear. To predict the model output power is an extremely important and difficult task without the complex machine learning algorithm. MGP performs very well to capture the original pattern of output power. In order to capture the intrinsic behaviors of the data, MGP plays a great role as it allows different types of turbines to capture information from other types of turbine in the cross-correlation structure. Sensitivity analysis also showed us the input-output nonlinear interactions. In future, a unified MGP model for the whole steam turbine power plant can be built such that we can have a total output power in the output end. Also, we can think of a depended GP structure such that we can build the model for the time series data not just the steady state values of the output. Also, this model can be employed if we can get the real data in the future.